\documentstyle[preprint,aps,epsfig]{revtex}

\begin{document}
\draft
\preprint{}
\def\qslash{\hbox{q\kern-.5em\lower.1ex\hbox{/}}}

%%%%%%%%%%%%%%%%%%%%%%%%%%%%%%%%%%%%%%%%%%%%%%%%%%
%\documentstyle[12pt,epsfig]{article}
%\textwidth 6.0in
%\textheight 8.6in
%\renewcommand{\baselinestretch}{1.5}
%\thispagestyle{empty}
%\topmargin -0.25truein
%\hoffset -.30in
%\flushbottom
%\parindent=1.5pc
%\baselineskip=24pt
%\begin{document}
%%%%%%%%%%%%%%%%%%%%%%%%%%%%%%%%%%%%%%%%%%%%%%%%%%%%

\begin{flushright}
USM-TH-80
\end{flushright}

\bigskip\bigskip
{\centerline{ \bf\huge Electric Screening Mass of the Gluon with
Gluon }} {\centerline{\bf\huge Condensate at Finite Temperature }}

\vspace{22pt}

\centerline{
 {Iv\'an Schmidt$^1$ and
Jian-Jun Yang $^{1,2}$}}

\vspace{8pt} {\centerline{\it $^1$ Departamento de F\'{\i}sica,
Universidad T\'ecnica Federico Santa Maria,}} {\centerline {\it
Casilla 110-V, Valpara\'{\i}so,  Chile}}

{\centerline{\it
$^2$ Department of Physics, Nanjing Normal  University,
Nanjing 210097, P. R. China}}

\vspace{10pt}
\begin{center}
{\large \bf Abstract}
\end{center}

The electric screening mass of the gluon at finite temperature is
estimated  by considering the gluon condensate above the critical
temperature. We find that the thermal gluons  acquire an electric
mass of order $T$ due to the gluon condensate.

\vspace{1cm}

\noindent PACS number(s): 12.38.Mh; 12.38.Lg; 52.25.Kn

\newpage

{\centerline {\bf 1. Introduction}}

\medskip

The color screening effect  is one of main features of the high
temperature quark-gluon plasma(QGP). The value of the electric
screening mass $(m_e\sim g(T)T)$, which gives rise to the Debye
screening of the heavy quark potential, has been known in
perturbative theory for quite some time\cite{Shuryak,Kalashnikov}.
Although much progress in perturbative
calculations\cite{Braaten90} has been achieved by the newly
developed techniques of resumed perturbation theory, it seems that
non-perturbative effects still dominate in the temperature regime
attainable in the near future heavy ion experiments. Especially in
the regime right  above the critical temperature, nonperturbative
effects are supposed to be important.

Quark and gluon condensates, which  describe the nonperturbative
effects of the QCD ground state, have been extensively used in QCD
sum rules\cite{SVZ} in order to study hadron properties  at zero
and finite temperature. Recently, the quark propagator at finite
temperature including the  gluon condensate was calculated by
Sch$\ddot{a}$fer and Thoma\cite{Thoma} and also the contribution
of the gluon condensate to the gluon propagator at zero
temperature has been extensively studied
\cite{JJYang,Lavelle1,Lavelle2,Lavelle3,Lavelle4,Lavelle5,Bagan,Elias}.
Here we will extend these investigations to the study of the self
energy of the gluon at finite temperature by using the thermal
gluon propagator.  The main purpose  of this paper aims at
estimating  the gluon contribution to the electric screening mass
of the gluon at finite temperature.

\vspace{0.5cm}

{\centerline {\bf 2. Gluon Condensate Contribution to Electric
Screening Mass }} {\centerline {\bf of the Gluon at Finite
Temperature}}

\medskip

The  lowest order contribution of the gluon, ghost and quark
condensates to the gluon propagator is depicted by diagrams in
Fig.~1. The presence of the  quark condensate comes from  chiral
symmetry breaking, so its  contributions in Fig.~1(e-f) vanish due
to chiral symmetry restoration  above the phase transition. The
gluon condensate above the critical temperature which is
associated with the breaking of  scale invariance has been
measured on the lattice recently\cite{Lattice}. We will focus on
the gluon condensate contribution to the self energy of the gluon.
In principle, in order to preserve the Slavnov-Taylor
identity(STI)\cite{STI}, the ghost condensate contributions in
Fig.1(b-c) should be taken into account to make sure that the
gluon self energy including the gluon condensate is transverse,
which makes the problem  complicated since  the expression for the
gluon self energy contains unknown condensates. However, as
pointed by Lavelle \cite{Lavelle4}, the gluon condensate is always
the main contribution to the nonperturbative gluon  at $T=0$. We
assume that this still holds in the  case  $T\not= 0$. To avoid
unnecessary complications, here we only analyze the gluon
condensate  contribution to the self energy of the gluon at small
momenta and extract the electric screening mass of the gluon due
to nonperturbative effects above the phase transition.

 The real part of the gluon self-energy in the one-loop
approximation is determined by the 1-1 component of the Feynman
propagator in thermo-field dynamics (TFD). Similar to the
perturbative case, using the Feynman rules of thermo-field
dynamics \cite{FR1,FR2} and adopting the imaginary time formalism,
the self-energy of the gluon with  the gluon condensate
contributions  shown in Fig.~1, is

\begin{eqnarray}
\Pi_{\mu\nu}(p_0,\vec{p})&=& -\frac{iN_c}{(N_c^2-1)}4 \pi \alpha_s
 i T \sum\limits_{k_0=2\pi i n T}  \int \frac{d^3k}{(2\pi)^3}
D^{(\rm{NP})}_{\lambda \lambda^\prime}(k_0,\vec{k})\nonumber \\
 &\times&[(2p-k)_\lambda g_{\mu\rho}+(-p+2k)_{\mu}g_{\rho
\lambda} +(-k-p)_{\rho}g_{\lambda\mu}]\nonumber \\
&\times&[(2p-k)_{\lambda^\prime}g_{\sigma \nu}
+(-p-k)_{\sigma}g_{\nu \lambda^\prime}
+(2k-p)_{\nu}g_{\lambda^\prime \sigma}] \nonumber\\ &\times&
[-\frac{g^{\rho\sigma}}{(p-k)^2}+\frac{(1-\xi)(p-k)^{\rho}
(p-k)^\sigma}{(p-k)^4}] \label{PI}
\end{eqnarray}
where at finite temperature the zeroth components of momentum
4-vectors take on discrete values, namely, $k_0=2 \pi i n T$ with
integer n. This is a direct consequence of Fourier analysis in the
imaginary time formation of Matsubara\cite{Fetter}. The lowest
Matsubara mode with $n=0$ should be a good approximation as long
as $p$ is not much larger than the critical temperature $T_c$. At
zero temperature, the transversality and Lorentz-invariance of the
gluon polarization tensor requires it to have the form
$\Pi_{\mu\nu}(k)=\Pi(k^2)(g_{\mu\nu}-k_\mu k_\nu /k^2)$. At finite
temperature, Lorentz-invariance is lost and the polarization
operator presents a combination of transverse and longitudinal
tensor structures. Therefore, the nonperturbative gluon propagator
$ D^{(\rm{NP})}_{\lambda \lambda^\prime}=D^{full}_{\lambda
\lambda^\prime}-D^{pert}_{\lambda \lambda^\prime}$, which contains
the gluon condensate, is generally assumed to have the form
\cite{Kapusta89}

\begin{equation}
D^{(\rm{NP})}_{\lambda
\lambda^\prime}(k_0,\vec{k})=D_L(k_0,\vec{k})P^L_{\lambda
\lambda^\prime} +D_T(k_0,\vec{k})P^T_{\lambda
\lambda^\prime},\label{DNP}
\end{equation}
where $D_{T,L}$ are the transverse and longitudinal parts of the
nonperturbative gluon propagator at finite temperature, and  the
bare gluon propagator has been subtracted since we are not
interested in perturbative corrections to the gluon self
energy\cite{Thoma}. In (\ref{DNP}), the longitudinal and
transverse projectors can be written as

\begin{equation}
P^L_{\lambda \lambda^\prime}=\frac{k_{\lambda}
k_{\lambda^\prime}}{k^2}-g_{\lambda \lambda^\prime}-P^T_{\lambda
\lambda^\prime}
\end{equation}

\begin{equation}
P^T_{\lambda 0}=0, P^T_{ij}=\delta_{ij}-\frac{k_i k_j}{k^2}
\end{equation}

Using the same method as  in Ref. \cite{Thoma}, we expand the
gluon propagator  in (\ref{PI}) for small loop momenta k and keep
only terms which are bilinear in k to relate the gluon condensate
to moments of the gluon propagator, and we obtain

\begin{eqnarray}
\Pi_{00}(p_0=0,\vec{p})&=&  \frac{4\pi\alpha_s N_c}{(N_c^2-1)}  T
\int  \frac{d^3\vec{k}}{(2\pi)^3}(\frac{8}{9}D_T-
\frac{136}{45}D_L)\frac{\vec{k}^2}{\vec{p}^2}.
\end{eqnarray}
The moments of the longitudinal and transverse gluon propagators
are related to the chromoelectric and chromomagnetic condensates
via\cite {Thoma}

\begin{equation}
\langle \vec{E}^2 \rangle _T=8T \int \frac{d^3\vec{k}}{(2\pi)^3}
\vec{k}^2 D_L (0,\vec{k}),
\end{equation}

\begin{equation}
\langle \vec{B}^2 \rangle _T=-16T \int \frac{d^3\vec{k}}{(2\pi)^3}
\vec{k}^2 D_T (0,\vec{k}).
\end{equation}
Therefore, $\Pi_{00}(p_0=0,\vec{p})$ can be re-written as

\begin{eqnarray}
\Pi_{00}(p_0=0,\vec{p}) &=& \frac{4\pi\alpha_s N_c}{(N_c^2-1)}
[-\frac{1}{18}\langle \vec{B^2}\rangle_T-\frac{17}{45}\langle
\vec{E^2}\rangle_T]\frac{1}{\vec{p}^2}.
\end{eqnarray}
Furthermore, from the expectation values of the space and timelike
plaquettes $\Delta_{\sigma,\tau}$ of lattice QCD\cite{Lattice},
the chromoelectric and chromomagnetic condensates can be extracted
as

\begin{equation}
\frac{\alpha_s}{\pi}\langle \vec{E}^2\rangle
_T=\frac{4}{11}\Delta_\tau T^4-\frac{2}{11}\langle
\vec{G}^2\rangle_{T=0},
\end{equation}

\begin{equation}
\frac{\alpha_s}{\pi}\langle \vec{B}^2\rangle
_T=-\frac{4}{11}\Delta_\sigma T^4+\frac{2}{11}\langle
\vec{G}^2\rangle_{T=0},
\end{equation}
Hence, we obtain,

\begin{eqnarray}
\Pi_{00}(p_0=0,\vec{p})
 &=&
\frac{4N_c\pi^2}{(N_c^2-1)\vec{p}^2} [\frac{2}{99} (\Delta_\sigma-
\frac{34}{5}\Delta_\tau)T^4 + \frac{29}{495} \langle G^2
\rangle_{T=0}]
\end{eqnarray}
where the gluon condensate at zero temperature is taken
as\cite{Lattice}

\begin{equation}
\langle G^2\rangle _{T=0} =(2.5\pm1.0)T_c^4
\end{equation}
The critical temperature is taken as  $T_c=260 MeV$\cite{Lattice}
in the following calculations.

The electric screening mass is related to the low momentum
behavior of the gluon polarization tensor,
 $\Pi_{\mu\nu}(p_0,\vec{p})$. It is generally  defined as the zero
momentum limit $(|\vec{p}|\to 0)$ in the static sector $(p_0=0)$
of $\Pi_{00}(p_0,\vec{p})$ \cite{Kapusta89}, i.e.,

\begin{equation}
m_e^2=\Pi_{00}(0,|\vec{p}| \to 0)
\end{equation}
 However, the above definition cannot be  correct  since
beyond leading order in the coupling it is gauge dependent in
non-Abelian theories\cite{Rebhan,Rebhan93}. A better way is to
define the electric  screening  mass as the position of the pole
of the propagator at spacelike momentum\cite{Rebhan,Rebhan93}

\begin{equation}
p^2+\Pi_{00}(0,p^2= -m_e^2)=0,\label{MD}
\end{equation}
where $p=|\vec{p}|$. Using this definition, the leading
perturbative contribution to the electric screening mass
reads\cite{Shuryak},

\begin{equation}
m_e^2(T)=(\frac{N_c}{3}+\frac{N_f}{6})g^2(T)T^2.
\end{equation}

Using the lattice results for the plaquette expectation
values\cite{Lattice}, the electric screening mass can be solved
from Eq.~(\ref{MD}). The numerical results are shown in
Fig.~\ref{plb1f2}. From Fig.~\ref{plb1f2}, one can  see  that the
screening mass due to the gluon condensate is almost proportional
to T when $T>1.3T_c$. In addition, the electric screening mass is
not sensitive to the value of the zero temperature gluon
condensate.

{\centerline {\bf 3. Summary and Discussion }}

To sum up, we investigated the gluon condensate contribution to
the electric screening mass of the gluon at finite temperature. In
the plasma, the gluons acquire an electric screening mass which is
approximately  proportional to  $T$. This  screening mass comes
mainly from the thermal gluon condensate and is not sensitive to
the value  of the zero temperature gluon condensate. The electric
mass gives rise to the Debye screening of the heavy quark
potential. The magnitude of $m_e$ influences strongly the
existence or non-existence of charmonium in the high temperature
phase according the following behaviour of the potential $V(r)$
between gauge invariant sources\cite{Michel}:

\begin{equation}
V(r) \propto \frac{\rm{e}^{-2 m_e r}}{r^2}.
\end{equation}
Therefore, the gluon screening mass of order $T$ due to
nonperturbative effects is supposed to influence the dissociation
of charmonium in the QGP.

In principle, this work can be extended to study the screening
behavior of the gluon magnetic sector. It is well known that the
absence of static magnetic screening in the hard thermal loop
resummed propagator leads to infrared singularities in
perturbative calculations, i.e. the zero value of the magnetic
mass-squared results in the divergence of the expansion of the
thermodynamic potential. However, a static magnetic screening
always vanishes in  perturbative calculations. A magnetic mass of
the gluon might appear in the nonperturbative gluon propagator.
Unfortunately, in order to study the transversality of the gluon
self energy, the STI should be considered exactly. Hence the ghost
condensate and higher order condensates at finite temperature have
to be included.

\begin{center}
{\bf ACKNOWLEDGEMENTS}
\end{center}
We would like to thank Drs. M. Lavelle and M.H. Thoma for
discussions. This work was supported in part by Fondecyt (Chile)
postdoctoral fellowship 3990048, by Fondecyt (Chile) grant 1990806
and by a C\'atedra Presidencial (Chile), and also by Natural
Science Foundation of China grant 19875024.

\vspace{0.5cm}
\begin{figure}[htb]
\begin{center}
\leavevmode {\epsfysize=5cm \epsffile{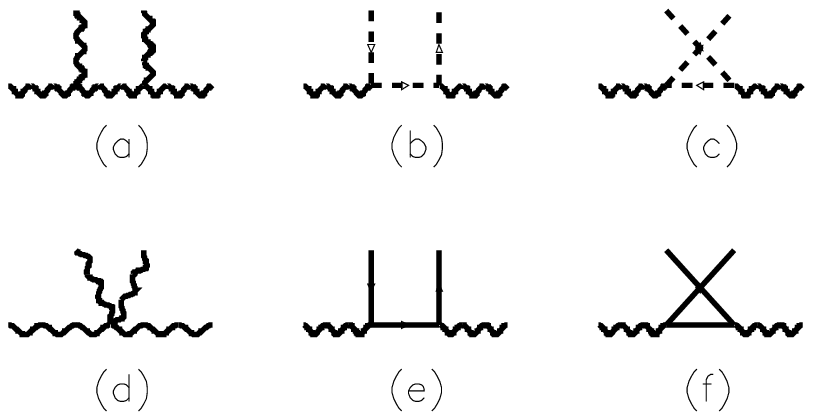}}
\end{center}
%\caption{}
\caption[*]{\baselineskip 13pt The Feynman diagrams for the
contributions of  the nonperturbative corrections to the gluon
propagator  with the lowest dimensional gluon, ghost and quark
condensates.} \label{plb1f1}
\end{figure}

\vspace{0.5cm}
\begin{figure}[htb]
\begin{center}
\leavevmode {\epsfysize=5cm \epsffile{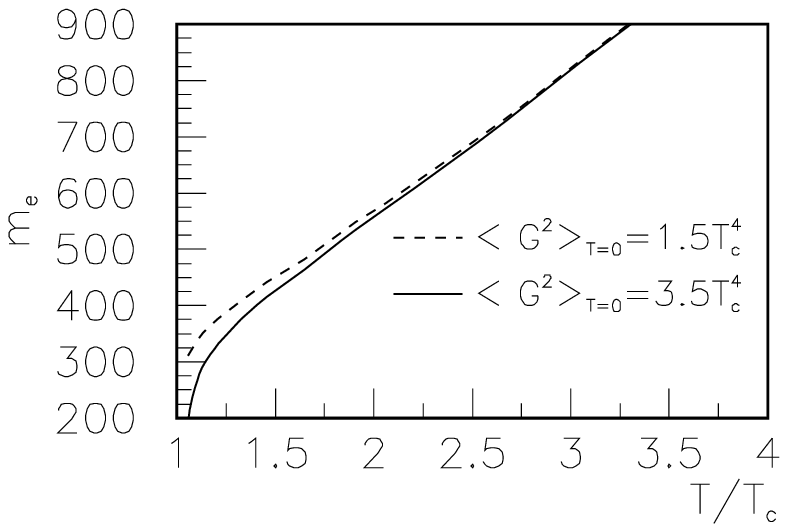}}
\end{center}
%\caption{}
\caption[*]{\baselineskip 13pt The electric screening mass $m_e$
versus $T/T_c$} \label{plb1f2}
\end{figure}


\begin{thebibliography}{150}

\bibitem{Shuryak}
E.V. Shuryak, Sov. Phys. JETP {\bf {47}}, 212 (1978) .




\bibitem{Kalashnikov}
O.K. Kalashnikov, Fortschr. Phys. {\bf{32}}, 525 (1984).


\bibitem {Braaten90}
E. Braaten and R. Pisarski,  Nucl. Phys.  {\bf{B337}}, 569 (1990).


%4
\bibitem {SVZ}
M. A. Shifman, A. I. Vainshtein and V. I. Zakharov, Nucl. Phys.
{\bf{B147}}, 385, 448, 519 (1979); J. Schwinger, {\sl Quantum
Chromodynamics,} Chap. IV., Eds. W. Beiglbock et al.,
(Spring-Velag, New York, 1983).



%1
\bibitem {Thoma}
A. Sch$\ddot{a}$fer and M. H. Thoma, Phys. Lett.  {\bf{B 451}},
195 (1999).



%9
\bibitem {JJYang}
J.J. Yang, H.Q. Shen, G.L. Li, T. Huang and P.N. Shen Nucl. Phys.
{\bf A640}, 457 (1998).


%10
\bibitem {Lavelle1}
M.J. Lavelle and M. Schaden, Phys. Lett. B {\bf{208}}, 297 (1988).


%11
\bibitem{Lavelle2}
M. Lavelle and M. Schaden, Phys. Lett. B {\bf{217}}, 551 (1989).



%12
\bibitem{Lavelle3}
M. Lavelle and M. Schaden, Phys. Lett. B {\bf{246}}, 487 (1990).

%13
\bibitem{Lavelle4}
M. Lavelle, Phys. Rev. D {\bf{44}}, R26 (1991).



%14
\bibitem{Lavelle5}
J. Ahlbach, M. Lavelle, M. Schaden and A. Streibl, Phys. Lett. B
{\bf{275}}, 124 (1992).




%15
\bibitem{Bagan}
E. Bagan and T. G. Steele, Phys. Lett. B {\bf{219}}, 497 (1989);
E. Bagan, M. R. Ahmady, V. Elias and T. G. Steele, Z. Phys. C
{\bf{61}}, 157 (1994).




%16
\bibitem {Elias}
V. Elias, T. G. Steele and M. D. Scadron, Phys. Rev. D {\bf{25}},
1584 (1988).



\bibitem{Lattice}
G. Boyd et al., Nucl. Phys. {\bf{B469}}, 419 (1996).






%21
\bibitem {STI}
J. C. Taylor, Nucl. Phys. {\bf{B33}}, 436 (1971); A. A. Slavnov,
Theor. Math. Phys. {\bf{10}}, 99 (1972).






\bibitem{FR1} N.P. Landsman and Ch. G. Van Weert, Phys. Rep.
{\bf{145}}, 141 (1987).

\bibitem{FR2} R.L. Kobes and G.W. Semenoff, Nucl. Phys.
{\bf{B 260}}, 714 (1985).


\bibitem{Fetter}
A.L. Fetter and J.D. Walecka, Quantum Theory of Many-Particle
Systems. McGraw-Hill. San Francisco, 1971.

\bibitem{Kapusta89}
J.I. Kapusta, Finite Temperature Field Theory, Cambridge
University press, Cambridge 1989.





\bibitem {Rebhan}
A.K. Rebhan, Nucl. Phys. {\bf{B430}}, 319 (1994).


\bibitem {Rebhan93}
A.K. Rebhan,  Phys. Rev.  {\bf{D48}}, R3967 (1993).

\bibitem {Michel}
M.L. Bellac, Thermal Field Theory, Cambridge University Press,
Cambridge 1996.




\end{thebibliography}
\end{document}